\documentstyle[editedvolume,epsf]{crckapb} 
\begin{opening}
\title{SURVEY OF FINE-SCALE STRUCTURE \protect\\
	 IN THE FAR-INFRARED MILKY WAY$^1$}

\author{WILLIAM H. WALLER \& FRANK VAROSI}
\institute{Hughes STX \& StarStuff Inc., NASA Goddard Space Flight Center, 
Code 681, Greenbelt, MD 20771}
\author{FRANCOIS BOULANGER}
\institute{Institut d'Astrophysique Spatiale, Universit\'e Paris XI, 
Batiment 121, 91405 Orsay Cedex, France}
\author{SETH W. DIGEL}
\institute{Hughes STX, NASA Goddard Space Flight Center}
\end{opening}
\runningtitle{Far-Infrared Milky Way}
\begin{document}
\section{MAPPING THE GALACTIC ISM}

\noindent\underbar {MOTIVATING QUESTIONS:}  
What is the general morphology of the diffuse
interstellar medium?  Is it mostly uniform or clumpy?  Are the clumps mostly
in the form of spheroidal clouds, sinuous filaments, extended sheets,
or discrete shells?  And do the {\it clumps} or the {\it voids} 
better define the overall 
structure?
By addressing these morphological questions, one can better constrain 
the 
dynamical processes that are most responsible for shaping and energizing the 
ISM.  
\vskip 6pt

\noindent\underbar{ANALYTIC STRATEGIES:}  By mapping the FIR emission from dust 
that has been warmed by the interstellar radiation field (ISRF), one can trace 
both the cool and warm phases of the diffuse ISM.  These two phases represent 
most of the mass in the diffuse ISM.  Recent data products produced by IPAC 
from the IRAS mission database provide the best resolved and most complete 
mapping of the Galactic FIR emission.  
And 
through spatial filtering, the strong gradient in brightness towards the 
Galactic midplane can be eliminated, thereby revealing the fine-scale FIR 
structure throughout the Galaxy.  
\vskip 6pt

\hang {\it $^1$This research was supported in part by a 
NASA Astrophysics Data Program 
contract (\#NAS5-32591) to StarStuff Incorporated.}

%
\begin{figure}
  \epsfysize=4.8cm
  \centerline{\vbox{\epsfbox{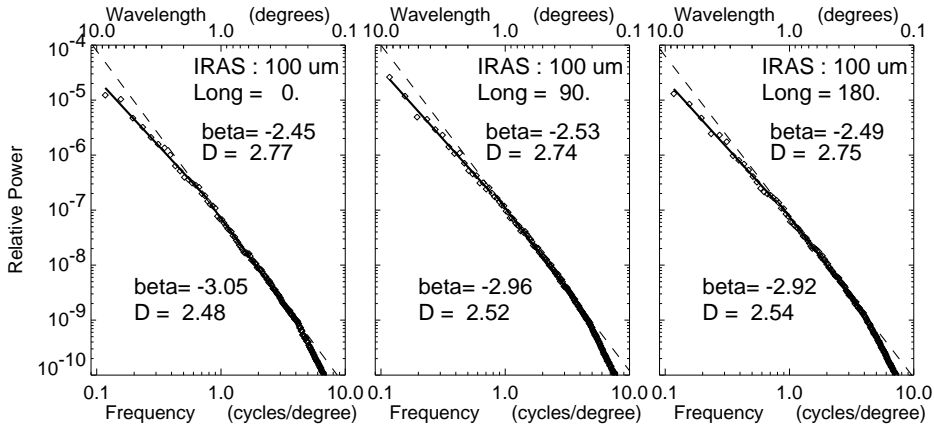}}}
\caption{Spatial power spectra of the FIR ``froth.'' }
\end{figure}

\section {REVEALING STRUCTURE IN THE FIR MILKY WAY}

We have produced a 
$360^{\circ} \times 60^{\circ}$ mural of 100$\mu$m emission in the Milky
Way from $60^{\circ} \times 60^{\circ}$ mosaics.  These mosaics
were
made from the IRAS Infrared Sky Atlas ``plates'' using the {\it SkyView}
Virtual Observatory (URL = http://skview.gsfc.nasa.gov/skyview.html).  
By applying a median normalizing spatial filter, we were able to 
eliminate the strong gradient in brightness towards the Galactic midplane.  
The resulting images reveal a ``froth'' of superposed filaments, voids, and 
shells (see {\bf Plate 1}).  

This fine-scale structure extends all the way down to 
the Galactic midplane.  Moreover, it scales in intensity with the smoothly 
varying background, independent of latitude, thus suggesting that the 
fine-scale residual emission is co-extensive with the smooth background.  {\it 
We conclude that the fine-scale structure is not merely of local origin, but 
consists of both nearby and more distant features in the disk}.

Although we had expected to find morphological evidence for supernova-driven 
``worms'' or ``chimneys'' rooted in the Galactic plane, our processing shows 
the FIR fine-scale structure to be more complex (e.g. less coherent and less 
rooted) as viewed in projection.  The observed FIR ``froth'' 
is just beginning to be identified with other 
tracers of the interstellar medium (e.g. CO, HI, and radio-continuum --- {\it 
cf. Wall and Waller, these Proceedings}).

Analysis of the spatial statistics shows that the FIR fine-scale structure is 
self-similar with a angular power-law exponent of $\beta \approx -3$ and and a 
fractal dimension of $D \approx 2.5$ --- similar to that found in 
isolated cirrus and molecular clouds.  On scales larger than $1.5^{\circ}$, the 
power-law exponent flattens to $\beta \approx -2.5$, perhaps indicating a 
change in the characteristic structure (see {\bf Figure 1}).  
This could be due to different 
dynamical inputs organizing the small and large-scale structures (e.g. 
turbulence and diffusion on small scales vs. macroscopic winds and shock fronts 
on larger scales). 
%
%

\begin{figure}
\caption{ {\bf (Plate 1)} Fine-scale structure in the 100 $\mu$m
emission for the $60^{\circ} \times 60^{\circ}$ field centered at
$(\ell,b) = (100^{\circ},0^{\circ})$.  Prominent emitting structures
include the Cygnus star-forming region at $(G78 + 2)$, IC 1396 at $(G99
+ 4)$ which seems to form part of a giant shell, NGC 7822 at $(G118 +
6)$ which also shows a shell-like morphology, and the galaxy M31 at
$(G121 - 21)$.  The remaining fine-scale features probably represent a
superposed mix of nearby and more distant filaments, voids, and shells
(see also Wall \& Waller, these Proceedings).}
\end{figure}

%
\end{document}